\documentclass[a4paper,twocolumn,11pt,accepted=2021-06-08]{quantumarticle}
\pdfoutput=1
\usepackage[utf8]{inputenc}
\usepackage[english]{babel}
\usepackage[T1]{fontenc}
\usepackage{amsmath}
\usepackage{hyperref}

\usepackage{tikz}
\usepackage{lipsum}
\usepackage{adjustbox}

\usepackage[numbers,sort&compress]{natbib}

\begin{document}

\title{Variational quantum solver employing the PDS energy functional}

\author{Bo Peng}
\email{peng398@pnnl.gov}
\affiliation{Physical and Computational Science Division, Pacific Northwest National Laboratory, Richland, Washington 99354, United States of America}
\orcid{0000-0002-4226-7294}
\author{Karol Kowalski}
\email{karol.kowalski@pnnl.gov}
\orcid{0000-0001-6357-785X}
\affiliation{Physical and Computational Science Division, Pacific Northwest National Laboratory, Richland, Washington 99354, United States of America}
\maketitle

\begin{abstract}
Recently a new class of quantum algorithms that are based on the quantum computation of the connected moment expansion has been reported to find the ground and excited state energies. In particular, the Peeters-Devreese-Soldatov (PDS) formulation is found variational and bearing the potential for further combining with the existing variational quantum infrastructure. Here we find that the PDS formulation can be considered as a new energy functional of which the PDS energy gradient can be employed in a conventional variational quantum solver. In comparison with the usual variational quantum eigensolver (VQE) and the original static PDS approach, this new variational quantum solver offers an effective approach to navigate the dynamics to be free from getting trapped in the local minima that refer to different states, and achieve high accuracy at finding the ground state and its energy through the rotation of the trial wave function of modest quality, thus improves the accuracy and efficiency of the quantum simulation. We demonstrate the performance of the proposed variational quantum solver for toy models, H$_2$ molecule, and strongly correlated planar H$_4$ system in some challenging situations. In  all  the case studies, the proposed variational quantum approach outperforms the usual VQE and static PDS calculations even at the lowest order. We also discuss the limitations of the proposed approach and its preliminary execution for model Hamiltonian on the NISQ device.
\end{abstract}

\section{Introduction}

Quantum computing (QC) techniques attract much attention in many mathematics, physics, and chemistry areas by providing means to address insurmountable computational barriers for simulating quantum systems on classical computers.\cite{nielsen2002quantum,shor1999polynomial,preskill2018quantum,low_depth_Chan,GoogleHF2020,mcardle2020quantum} 
One of the focus areas for quantum computing is quantum chemistry, where Hamiltonians can be effectively mapped into qubit registers. In this area, several quantum computing algorithms, including quantum phase estimator (QPE)
\cite{luis1996optimum, cleve1998quantum,berry2007efficient,childs2010relationship,seeley12_224109,
PhysRevA.92.042303,haner2016high,poulin2017fast}
and variational quantum eigensolver (VQE),
\cite{peruzzo2014variational,mcclean2016theory,fontalvo2017strategies,PhysRevA.95.020501,kandala2017hardware,kandala2018extending,grimsley2019adaptive,PhysRevX.8.011021,huggins2020non}
have been extensively tested on benchmark systems corresponding to the description of chemical reactions involving bond-forming and breaking processes, excited states, and strongly correlated molecular systems. In more recent applications, several groups reported quantum algorithms for imaginary time evolution,\cite{mcardle2019variational,motta2020determining} 
quantum filter diagonalization,\cite{parrish2019quantum} quantum inverse iteration algorithms,\cite{kyriienko2020quantum} and quantum power/moments methods. \cite{Seki2021quantum,Vallury2020quantumcomputed} 
The main thrust that drives this field is related to the efficient encoding of the electron correlation effects that are needed to describe molecular systems.

Basic methodological questions related to an efficient way of incorporating large degrees of freedom required to capture a subtle balance between static and dynamical correlations effects still need to be appropriately addressed. A typical way of addressing these challenges in VQE approaches is by incorporating more and more parameters (usually corresponding to excitation amplitudes in a broad class of unitary coupled-cluster methods \cite{hoffmann1988unitary,unitary1,unitary2,kutzelnigg1991error,lee2018generalized,evangelista2019exact}). Unfortunately, this brute force approach is quickly stumbling into insurmountable problems associated with the resulting quantum circuit complexity and problems with numerical optimization procedures performed on classical machines (the so-called barren plateau problem reported in  
Refs.\cite{mcclean2018barren,cerezo2020cost,wang2020noise,cerezo2020impact,pesah2020absence,marrero2020entanglement,uvarov2020barren}).

In this paper, we propose a new solution to these problems. Instead of adding more parameters to the trial wave function, we choose to optimize a new class of energy functionals (or quasi-functionals, where the energy is calculated as a simple equation solution) that already encompasses information about high-order static and dynamical correlation effects. An ideal choice for such high-level functional is based on the Peeters, Devreese, and Soldatov (PDS) formalism,\cite{peeters1984upper,soldatov1995generalized} where variational energy is obtained as a solution of simple equations expressed in terms of the Hamiltonian’s moments or expectations values of the powers of the Hamiltonians operator defined for the trial wave function. In Ref. \cite{kowalski2020cmx} we demonstrated that in such calculations high-level of accuracy can be achieved even with very simple parametrization of the trial wave functions (capturing only essential correlation effects) and low-rank moments. We believe that merging the PDS formalism with the quantum gradient based variational approach would be considered as a more interesting alternative for by-passing main problems associated with the excessive number of amplitudes that need to be included to reach the so-called chemical accuracy.

In the following sections we will 
\textcolor{black}{briefly introduce the PDS formalism}
and describe how the PDS energy functional can be incorporated with the minimization procedures that are based on the quantum gradient approach \cite{guerreschi2017practical,schuld2019gradient,mcclean2018barren,mcardle2019variational,yamamoto2019natural} to produce a new class of variational quantum solver (which is called PDS($K$)-VQS for short in the rest of the paper) to target the ground state and its energy in a quantum-classical hybrid manner. Furthermore, we will test its performance, in particular the performance of the more affordable lower order PDS($K$)-VQS ($K=2,3,4$) approaches combining with the trial wave function expressed in low-depth quantum circuits, at finding the ground state and its energy for the Hamiltonians describing toy models and H$_2$ molecular system, as well as the strongly correlated planar H$_4$ system, in some challenging situations where the barren plateau problem precludes the effective utilization of the standard VQE approach.

\section{Method}

\subsection{\textcolor{black}{PDS formalism}}

\textcolor{black}{In this section we will give a brief description of the PDS formalism. The detailed discussion of the PDS methodology and highly relevant connected moment expansion (CMX) formalisms have been given in the original work\cite{peeters1984upper,soldatov1995generalized} as well as our recent work\cite{kowalski2020cmx,claudino2021} and many earlier literatures (see for example Refs. \cite{knowles1987validity,prie1994relation,mancini1994analytic,ullah1995removal,mancini1995avoidance,fessatidis2006generalized,fessatidis2010analytic}).
The many-body techniques used in the derivation of PDS expansions originate in the effort to provide upper bounds for the free energies, and to provide alternative re-derivation of the Bogolubov's \cite{bogoliubov1947theory} and Feynman's \cite{feynman1955slow} inequalities. Since the Gibbs-Bogolubov inequality reduces to the Rayleight-Ritz variational principle in zero temperature limit, these formulations can be directly applied to quantum chemistry.  Here we only provide an overview of basic steps involved in the derivations of the PDS formulation.} 

\textcolor{black}{A starting point of the studies of upper bounds for the exact ground-state energy $E_0$ is the  analysis of function 
$\Gamma(t)$ (defined for trial wave function $|\phi\rangle$ having non-zero overlap with the ground-state wave function)
\begin{equation}
\Gamma(t)=\langle\phi|e^{-tH}|\phi\rangle \;,
\label{gam1}
\end{equation}
and its Laplace transform $f(s)$
\begin{equation}
    f(s)=\int_{0}^{+\infty} e^{-st} \Gamma(t) dt \;.
    \label{lap1}
\end{equation}
It can be proved that, for a complex scalar $s$, Eq. (\ref{lap1}) exists if the real part of $s$ $\Re (s) > - E_0$.
Under this condition, for Hamiltonian $H$ defined by discrete energy levels $E_i$
and corresponding eigenvectors $|\Psi_i\rangle$ ($i=0,1,\ldots,M$)
\begin{equation}
    H=\sum_{i=0}^{M} E_i |\Psi_i\rangle\langle\Psi_i|\;,
    \label{ham1}
\end{equation}
$f(s)$ takes the form 
\begin{equation}
    f(s)=\sum_{i=0}^{M} \frac{\omega(E_i)}{s+E_i}
    \label{fs}
\end{equation}
where $\omega(E_i)=|\langle\Psi_i|\phi\rangle|^2$. The PDS formalism is based on introducing  parameters into expansion (\ref{fs}) using  a simple identity (with a real parameter $a$)
\begin{widetext}
\begin{eqnarray} 
\frac{1}{s+E_n}
&=&\frac{1}{s+a}-\frac{E_n-a}{(s+a)^2} +
     \frac{(E_n-a)^2}{(s+E_n)(s+a)^2} \;.
\label{iden1}
\end{eqnarray}
\end{widetext}
When the above identity is applied for the first time to Eq. (\ref{fs}) (introducing the first parameter $a_1$) one gets the following expression for the $f(s)$ function 
\begin{widetext}
\begin{eqnarray}
    f(s)
    &=&\sum_{i=0}^{M} \omega(E_i)
    \left[
     \frac{1}{s+a_1}-\frac{E_i-a_1}{(s+a_1)^2} +
     \frac{(E_i-a_1)^2}{(s+E_i)(s+a_1)^2}
    \right]
    \label{step21}
\end{eqnarray}
\end{widetext}
The transformation (\ref{step21}) can be repeated $K$ times (with each time introducing a new parameter $a_i$, $i=1,\ldots,K$) to reformulate the $f(s)$ function as 
\begin{widetext}
\begin{equation}
f(s)=R_K(s,a_1,\ldots,a_K)+W_K(s,a_1,\ldots,a_K)\;,
\label{fsrnwn}
\end{equation}
\end{widetext}
where
\begin{widetext}
\begin{eqnarray}
R_K(s,a_1,\ldots,a_K) &=&  \sum_{i=0}^{M}
\left[\frac{\omega(E_i)}{s+E_i} \prod_{j=1}^K 
\frac{(E_i-a_j)^2}{(s+a_j)^2}\right] \ge - E_0 ~~(\text{if}~~ \Re(s) > -E_0), \label{xx1} \\
W_K(s,a_1,\ldots,a_K) &=& \sum_{i=0}^{M}
\left\{\omega(E_i)\sum_{j=1}^{K} \left[
\Big(
\frac{1}{s+a_j}-
\frac{E_i-a_j}{(s+a_j)^2}
\Big)
\prod_{n=1}^{j-1}
\frac{(E_i-a_n)^2}{(s+a_n)^2} \right] \right\}\;. \label{xx2}
\end{eqnarray}
\end{widetext}
The $K$-th order PDS formalism (PDS($K$) for short henceforth) is then associated with defining the introduced $K$ real parameters $(a_1,\ldots,a_K)$ that minimize the value of $R_K(s,a_1,\ldots,a_K)$. In this minimization process the necessary extreme conditions are given by the system of equations
\begin{equation}
    \frac{\partial R_K(s,a_1,\ldots,a_K)}{\partial a_i}=0,
     ~~(i=1,\ldots,K),
\end{equation}
which can be alternatively represented by the matrix system of equations for an auxiliary vector $\mathbf{X} = (X_1,\cdots,X_K)^T$
\begin{equation}
\mathbf{MX} = -\mathbf{Y}. \label{lineq1} 
\end{equation}
Here, the matrix elements of  $\bf M$ and vector $\bf Y$ are defined as the expectation values of Hamiltonian powers (i.e. moments), $M_{ij} = \langle\phi| H^{2K-i-j} |\phi \rangle$, $Y_i = \langle\phi| H^{2K-i} |\phi\rangle$ ($i,j = 1,\cdots,K$)  (for simplicity, we will use the notation $\langle H^n \rangle \equiv \langle\phi|H^n|\phi\rangle$). It can be shown that the optimal parameters in the PDS($K$) formalism, $(a_1^{(K)},\ldots,a_K^{(K)})$, are the roots of the polynomial $P_K(\mathcal{E})$,
\begin{equation}
P_K(\mathcal{E}) = \mathcal{E}^{K} + \sum_{i=1}^K X_i \mathcal{E}^{K-i}, \label{poly}   
\end{equation}
and these roots provide upper bounds for the exact ground and excited state energies, e.g., for the ground state energy we have
\begin{equation}
    E_0 \le {\rm min}(a_1^{(K)},\ldots,a_K^{(K)}) 
    \le \langle\phi|H|\phi\rangle \;.
\end{equation}
Note that, as shown in Refs.\cite{peeters1984upper,soldatov1995generalized}
the PDS formalism also applies to the Hamiltonian characterized by discrete and continuous spectral resolutions together.}

\subsection{PDS($K$)-VQS formalism}

In the variational method, we approximate the quantum state using parametrized trial state $|\Psi\rangle \approx |\phi\rangle$. Using a quantum circuit, the trial state can be prepared by applying a sequence of parametrized unitary gates on the initial state $|0\rangle$,
\begin{align}
|\phi\rangle = |\phi(\vec{\theta})\rangle = \cdots U_k(\theta_k) \cdots U_1(\theta_1) |0 \rangle
\end{align}
($\vec{\theta} = \{\theta_1,\cdots,\theta_n\}$). Here $U_k(\theta_k)$ is the $k$-th unitary single- or two-qubit gate that is controlled by parameter $\theta_k$. The goal is to approach the ground-state energy of a many-body Hamiltonian, $H$, by finding the values of these parameters, $\vec{\theta}$, that minimize the expectation value of the Hamiltonian
\begin{align}
E_{\min} = \min_{\vec{\theta}} \langle \phi(\vec{\theta})|H| \phi(\vec{\theta}) \rangle.
\end{align}

To do this, the conventional VQE starts by constructing the ansatz $|\phi(\vec{\theta}) \rangle$ and measuring the corresponding expectation value of the Hamiltonian using a quantum computer, and then relies on a classical optimization routine to obtain new $\vec{\theta}$. During the parameter optimization (or dynamics), the set of parameters that is updated at the $k$-th step ($k>1$) can be written as
\begin{align}
\vec{\theta}_{k} = \vec{\theta}_{k-1} - \eta \mathcal{R}^{-1}(\vec{\theta})\nabla \mathcal{E}(\vec{\theta}). \label{dynamics}
\end{align}
where $\nabla \mathcal{E}(\vec{\theta}) = \partial \mathcal{E}/\partial \vec{\theta}$ is the energy gradient vector, and $\eta$ is the step size (or learning rate).
$\mathcal{R}(\vec{\theta})$ is the Riemannian metric matrix at $\vec{\theta}$ that is flexible to characterize the singular point in the parameter space and is essentially related to the indistinguishability of $\mathcal{E}(\vec{\theta})$.\cite{yamamoto2019natural} 
It is worth mentioning that Eq. (5) originates from natural gradient learning method in the general nonlinear optimization framework especially targeting machine learning problems.\cite{amari1998NG} Here, the natural gradient is the optimizer that accounts for the geometric structure of the parameter space. For the curved (or nonorthonormal) parameter manifold that exhibits the Riemannian character (e.g. in large neural networks), natural gradient learning method is often employed to avoid the plateaus in the parameter space.\cite{mcardle2019variational, yamamoto2019natural,Stokes2020quantumnatural} 

Note that when the parameter space is a Euclidean space with orthonormal coordinate system the Riemannian metric tensor will reduce to the unity matrix (see Tab. \ref{gradient}).  In VQE setting, one can define the Riemannian metric as the quantum Fubini-Study metric, which is the quantum analog of the Fisher information matrix in the classical natural gradient,\cite{amari1998NG} to measure the distance in the space of pure quantum states. The quantum Fubini-Study metric describes the curvature of the ansatz class rather than the learning landscape, but often performs as well as Hessian based methods (e.g. BFGS optimizer that approximates the Hessian of the cost function using first-order gradient, see Ref. \cite{Wierichs2020NGD} for a recent detailed discussion). There are also some other options for the Riemannian metric including imaginary-time evolution (ITE) or even classical Fisher metric that have been discussed in some recent reports.\cite{mcardle2019variational, yamamoto2019natural,Stokes2020quantumnatural} In Tab. \ref{gradient}, three commonly used flavors of the Riemannian metric matrix $\mathcal{R}(\vec{\theta})$ are listed and will be used in the following case studies.
\textcolor{black}{Remarkably, as pointed out in Refs. \cite{Yuan2019theoryofvariational,mcardle2019variational}, the difference between natural gradient descent (NGD) and ITE accounts for the global phase, and if introducing a time-dependent phase gate to the trial state, the Riemannian metric employing NGD will be equivalent to the metric employing ITE.}

\begin{table}[]
\begin{adjustbox}{max width=\columnwidth}
\begin{tabular}{cc}
\hline \hline
\multicolumn{1}{c}{}& \multicolumn{1}{c}{\textbf{Riemannian metric} $\mathcal{R}_{ij}(\vec{\theta})$}  \\ \hline
& \\
\multicolumn{1}{c}{\textbf{GD}}&
\multicolumn{1}{c}{$\delta_{ij}$} \\ 
& \\
\multicolumn{1}{c}{\textbf{NGD}} &
\multicolumn{1}{c}{$\Re \Big( \frac{\partial \langle \phi(\vec{\theta})|}{\partial \theta_i} \frac{\partial | \phi(\vec{\theta})\rangle}{\partial \theta_j} \Big) - \frac{\partial \langle \phi(\vec{\theta})|}{\partial \theta_i} |\phi(\vec{\theta}) \rangle \langle \phi(\vec{\theta}) | \frac{\partial | \phi(\vec{\theta})\rangle}{\partial \theta_j}$} \\ 
& \\
\multicolumn{1}{c}{\textbf{ITE}}&
\multicolumn{1}{c}{$\Re \Big( \frac{\partial \langle \phi(\vec{\theta})|}{\partial \theta_i} \frac{\partial | \phi(\vec{\theta})\rangle}{\partial \theta_j} \Big)$} \\ 
& \\ \hline \hline
\end{tabular}
\end{adjustbox}
\caption{Three Rimannian metric forms, ordinary gradient descent (GD), natural gradient descent (NGD), and imaginary time evolution (ITE), exploited in the present study.}\label{gradient}
\end{table} 

\begin{figure}[h!]
    \includegraphics[width=\linewidth]{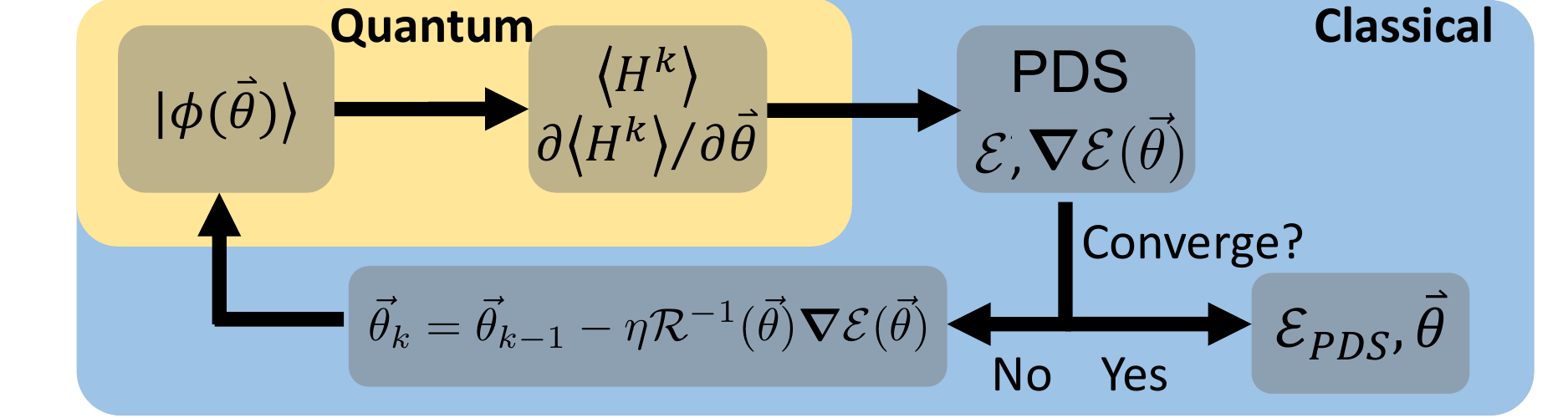}
    \caption{The workflow of variational quantum solver employing the PDS energy functional.}
    \label{workflow}
\end{figure}

To get the energy gradient in the PDS framework, take the derivative w.r.t. $\theta_i$ on both sides of Eq. (\ref{poly}), and after reorganizing the terms we can express the energy derivative as
\begin{widetext}
\begin{align}
\frac{\partial \mathcal{E}}{\partial \theta_i} = \frac{-1}{K\mathcal{E}^{K-1} + \displaystyle\sum_{i=1}^{K-1}(K-i)X_i\mathcal{E}^{K-i-1}} \left( 
\begin{array}{c}
\mathcal{E}^{K-1}  \\
\vdots \\
1
\end{array}\right)^T \frac{\partial \mathbf{X}}{\partial \theta_i}, \label{dE}
\end{align}
\end{widetext}
where $\frac{\partial \mathbf{X}}{\partial \theta_i}$ is associated with the $\theta_i$-derivative of Eq. (\ref{lineq1}),
\begin{align}
\mathbf{M} \frac{\partial \mathbf{X}}{\partial \theta_i} =
- \frac{\partial \mathbf{Y}}{\partial \theta_i} - \frac{\partial \mathbf{M}}{\partial \theta_i} \mathbf{X}, \label{lineq2}
\end{align}
and can be obtained by solving Eq. (\ref{lineq2}) as a linear equation
with $\partial Y_i/\partial \theta_k = \partial \langle H^{2K-i}\rangle/\partial \theta_k$ and $\partial M_{ij}/\partial \theta_k = \partial \langle H^{2K-i-j}\rangle/\partial \theta_k$.
Fig. \ref{workflow} summarizes the workflow of PDS($K$)-VQS, where on the classical side the PDS($K$) module includes two steps, (i) solving two consecutive linear problems to get $\mathbf{X}$ and $\partial\mathbf{X}/\partial\theta_i$, and (ii) solving for roots of polynomial (\ref{poly}) and computing Eq. (\ref{dE}). 
On the quantum side, in comparison with the conventional VQE, the present PDS($K$)-VQS infrastructure relies on quantum circuits to measure $\langle H^n \rangle$ and their $\vec{\theta}$-derivatives.

In the present work, due to the relatively small system size, we directly exploit the Hadamard test to compute the real part of $\langle H^n \rangle$ for the Hamiltonians that are represented as a sum of Pauli strings. It is worth mentioning that for typical molecular systems that can be represented by $N$ qubits, the number of $\langle H^n \rangle$ measurement scales as $\mathcal{O}(N^{4n})$, which nevertheless can be reduced once the Pauli strings are multiplied and their expectation values are re-used as the contributions to the higher order moments.\cite{kowalski2020cmx} 
For example, as we will show later for the H$_4$ system that comprises 184 Pauli strings in the Hamiltonian, the effective number of Pauli strings required for arbitrary $\langle H^n \rangle$ ($n=2,3,4$) measurements can be dropped from 184$^2$, 184$^3$, and 184$^4$ to 1774, 3702, and 4223, respectively, after the Pauli reduction, and the 4223 strings will not be changed for more complex $\langle H^n \rangle$'s ($n>4$). Similar findings have also been reported in Ref. \cite{Vallury2020quantumcomputed}, where by grouping the Pauli strings into tensor-product basis sets the authors examined the operator counts for $\langle H^4 \rangle$ of Heisenberg model defined on different lattice geometries for the number of qubits ranging from 2 up to 36, and found that the effective number of Pauli strings to be measured drops by several orders of magnitude with sub-linear scaling in a given number of qubits.
For larger systems, the number of measurements can be further reduced by introducing active space and local approximation. Alternatively, one can approximate $\langle H^n \rangle$ by a linear combination of the time-evolution operators as introduced in some recent reports.\cite{Seki2021quantum, bespalova2020hamiltonian} 
For the estimation of $\partial \langle H^n \rangle/\partial \theta_k$, in the present work we limit $U_k(\theta_k)$ exploited in the state preparation to be only one-qubit rotations. Then, following Ref. \cite{schuld2019gradient}, $\partial \langle H^n \rangle/\partial \theta_k$ can be obtained by measuring $\langle H^n \rangle$ twice using the same circuit but shifting $\theta_k$ by $\pm \frac{\pi}{2}$ separately, i.e.
\begin{widetext}
\begin{align}
\frac{\partial \langle H^n \rangle _{(\cdots,\theta_k,\cdots)} }{ \partial \theta_k} = \frac 1 2 \Big(
\langle H^n \rangle_{(\cdots,\theta_k+\frac{\pi}{2},\cdots)} - \langle H^n \rangle_{(\cdots,\theta_k-\frac{\pi}{2},\cdots)}
 \Big). \label{shift}
\end{align}
\end{widetext}
If $\theta_k$ parametrizes more than one one-qubit rotations in the circuit, then based on the product rule $\partial \langle H^n \rangle/\partial \theta_k$ will have contributions from all one-qubit $\theta_k$ rotations, each of which will be obtained by applying (\ref{shift}) on the corresponding rotation.

\section{Numerical examples}

In this section, with several examples, we will demonstrate how the PDS($K$)-VQS performs in some challenging situations, and its difference in comparison to the conventional VQE and static PDS($K$) expansions.

\subsection{Toy Hamoltonians}

We first test the PDS($K$)-VQS on two toy Hamiltonians
\begin{align}
H_A &= 1.5 I_{4\times4} + 0.5(I_{2\times 2}\otimes\sigma_z-2\sigma_z\otimes\sigma_z) \notag \\
&= \left(
\begin{array}{cccc}
1 & 0 & 0 & 0 \\
0 & 2 & 0 & 0 \\
0 & 0 & 3 & 0 \\
0 & 0 & 0 & 0 
\end{array} \right), \notag \\
H_B &= I_{4\times4} + 0.5(I_{2\times 2}\otimes\sigma_z-\sigma_z\otimes\sigma_z) \notag \\
&= \left(
\begin{array}{cccc}
1 & 0 & 0 & 0 \\
0 & 1 & 0 & 0 \\
0 & 0 & 2 & 0 \\
0 & 0 & 0 & 0 
\end{array} \right), \notag 
\end{align}
with ansatze
\begin{align}
&|\phi_A(\theta_1,\theta_2) \rangle 
= \tilde{R}_Y^{0,1}(\theta_2) R_X^0(\theta_1)|00\rangle, \notag \\
&|\phi_B(\theta_1,\theta_2) \rangle 
= \tilde{R}_Y^{0,1}(\theta_2) R_X^0(\theta_1) R_X^1(\theta_1) |01\rangle, \notag 
\end{align}
that have been exploited by McArdle et al. \cite{mcardle2019variational} to demonstrate the performance of different Riemannian metrics in the conventional VQE approach for finding the ground-state energy of the same Hamiltoinans. Here, $\tilde{R}_Y^{p,q}(\theta)$ is a controlled $Y$ rotation of $\theta$ with control qubit $p$ and target qubit $q$, and $R_X^p(\theta)$ is a rotation of $\theta$ on qubit $p$ around the $x$-axis. The rotation about the $j$-axis is defined as $R_{\sigma_j}(\theta) = e^{-\text{i}\theta\sigma_j/2}$ with $\sigma_j$ being one of the Pauli spin matrices. 

\begin{figure}[h!]
    \includegraphics[width=\linewidth]{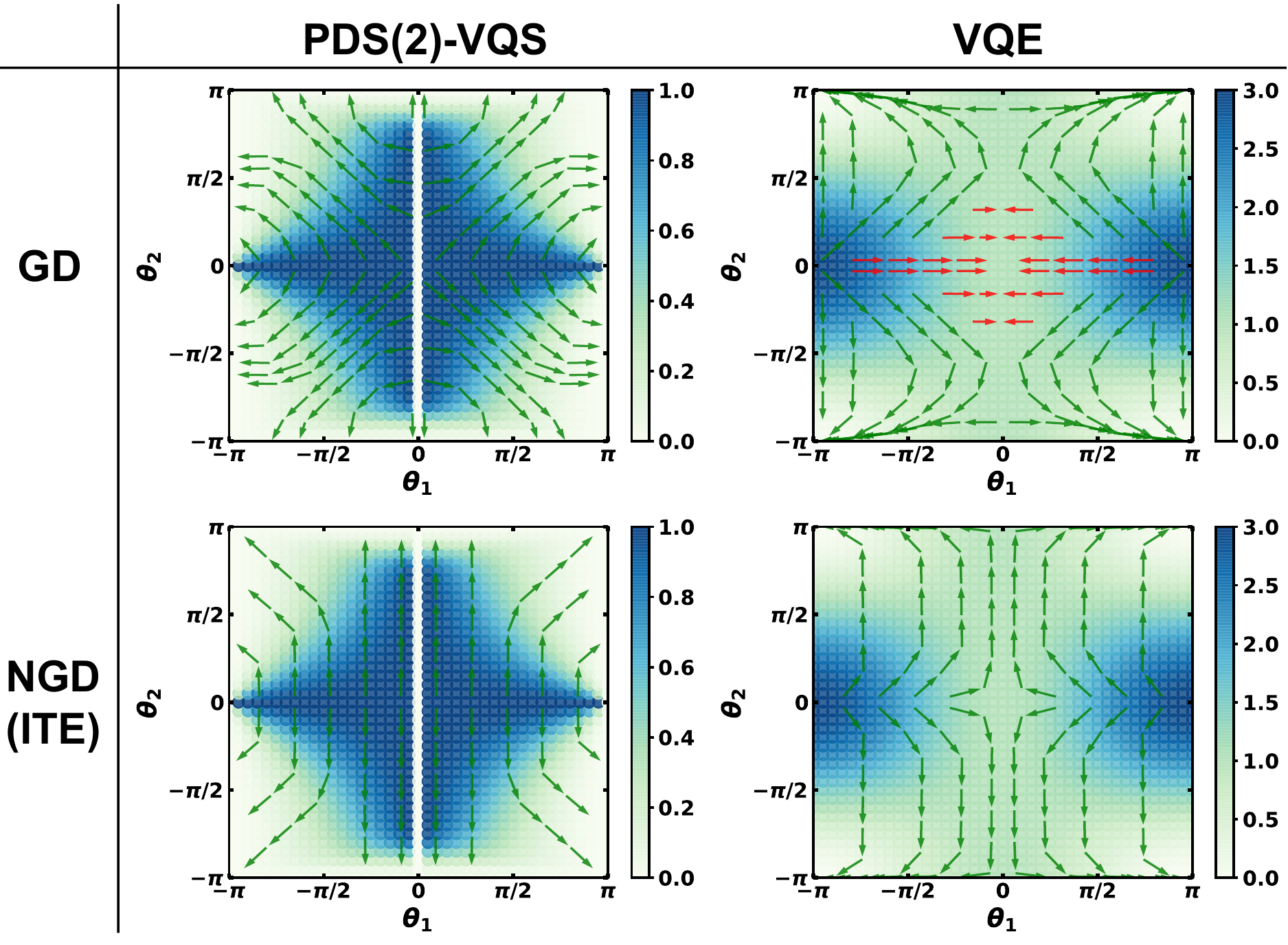}
    \caption{Variational trajectories on the PDS(2) energy surface (left panels) and original potential energy surface (right panels) discovering the ground state energy of Hamiltonian, $H_A$, explored by gradient descent (top panels) and natural gradient descent/imaginary time evolution (bottom panels). On the background energy surfaces, the dark blue and white colors correspond to the global maximum and minimum energies, respectively. The arrows indicate the trajectories of the dymanics, and are colored green if the trajectory converges to the ground state energy, and red otherwise. The step size $\eta=0.05$ in all the calculations.}
    \label{HA}
\end{figure}

\begin{figure}[h!]
    \includegraphics[width=\linewidth]{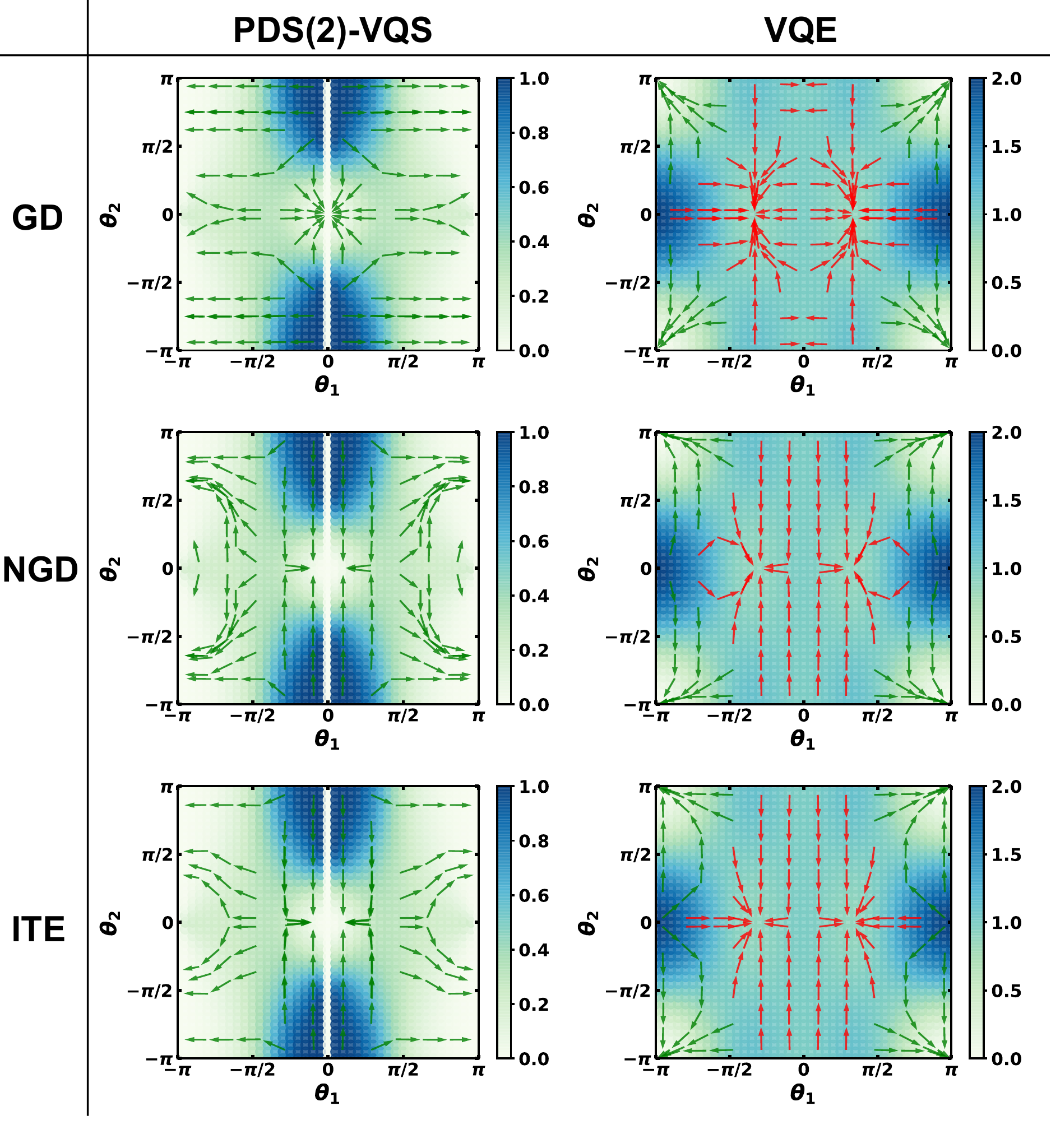}
    \caption{Variational trajectories on the PDS(2) energy surface (left panels) and original potential energy surfaces (right panels) discovering the ground state energy of Hamiltonian, $H_B$, explored by gradient descent (top panels), natural gradient descent (middle panels), and variational imaginary time (bottom panels). On the background energy surfaces, the dark blue and white colors correspond to the global maximum and minimum energies, respectively. The arrows indicate the trajectories of the methods, and are colored green if the trajectory converges to the true ground state energy, and red otherwise. The step size $\eta=0.05$ in all the calculationss.}
    \label{HB}
\end{figure}

Figs. \ref{HA} and \ref{HB} show the performances of the proposed PDS($K$)-VQS ($K=2$, i.e. PDS(2)-VQS) and the conventional VQE approaches for finding the ground state energy of the toy Hamiltonians. As can be seen, the ability of VQE navigation to avoid the local minima on the conventional PES depends on the Riemannian metric exploited. For system A, in comparison to GD, the NGD (or equivalently ITE in this case) is able to avoid the local minimum at $(\theta_1,\theta_2)=(0,0)$. This is because the Riemannian metric,
\begin{align}
\mathcal{R} = \left(
\begin{array}{cc}
\frac 1 4    & 0 \\
0 & \frac 1 4\sin^2(\frac{\theta_1}{2})
\end{array}
\right), \notag
\end{align}
used in the NGD/ITE correctly characterizes any rotation pair with $\theta_1=0$ as a singular point (i.e. $\det|R|=0$) such that $\mathcal{R}^{-1}$ will numerically navigate the dynamics (e.g. via singular value decomposition) to avoid collapsing in this local minimum once the trajectory is getting close. Therefore, if the metric is unable to characterize the local minima as singular points, the VQE would still get trapped. This can be observed from the VQE performance for system B, where both NGD and ITE fail to escape the local minima, $(\theta_1,\theta_2) \sim (\pm\frac{3\pi}{8},0)$, in the dynamics due to the fact that the local minima are not the singular points of $R$ in either NGD or ITE. 

In contrast, the PDS(2)-VQS robustly converge to the true ground state for both systems regardless of the employed Riemannian metric. The success of PDS($K$)-VQS in these toy examples can be essentially attributed to the fact that, in comparison to the original PES where the local minima corresponding to a non-ground state, the entire PDS($K$) energy surface, except the singular areas (see the infinitesimal white strips on the left panels of Fig. \ref{HA} at $\theta_1=0$) where the fidelity of the trial wave function w.r.t the target state is strictly zero, provides an upper bound energy surface for the same (ground) state. This state-specific nature makes the PDS($K$)-VQS essentially explore a lower upper bound of the ground state at a given PDS order, and therefore the dynamics will not be trapped at a location that is associated with a different state. It worth mentioning that, a lower bound of the ground state energy can also be obtained from a static, and more costly, higher order PDS($K$) standalone calculation as demonstrated in our previous work.\cite{kowalski2020cmx} From this perspective, the PDS($K$)-VQS approach provides an effective way to explore the possibility of pushing the low order PDS($K$) results towards high accuracy that would otherwise require higher-order and more expensive PDS($K$) calculations. Besides, since generalized variational principle applies in the PDS framework,\cite{peeters1984upper,soldatov1995generalized} if other roots of Eq. (\ref{poly}) are concerned, the PDS($K$)-VQS will also be able to navigate the dynamics to give lower upper bounds for excited states as long as the fidelity of the trial wave function with respect to the target state is non-zero. 

\subsection{$H_2$ and $H_4$ systems}

We further employ the proposed PDS($K$)-VQS approach to find the ground state energy of H$_2$ and H$_4$ molecular systems. For H$_2$ molecule, we exploit an effective Hamiltonian and an ansatz exploited by Yamamoto \cite{yamamoto2019natural} and Bravyi et al.\cite{bravyi2017tapering},
\begin{align}
&H = 0.4(\sigma_z \otimes I + I \otimes \sigma_z) + 0.2\sigma_x\otimes \sigma_x \notag \\
&|\phi(\vec{\theta})\rangle = R_Y^0(2\theta_3)R_Y^1(2\theta_4)\tilde{U}_N^{0,1}R_Y^0(2\theta_1)R_Y^1(2\theta_2)|00\rangle \notag
\end{align}
where $\tilde{U}_N^{p,q}$ denotes the CNOT gate with control qubit $p$ and target qubit $q$. 
\begin{figure}[h!]
    \includegraphics[width=\linewidth]{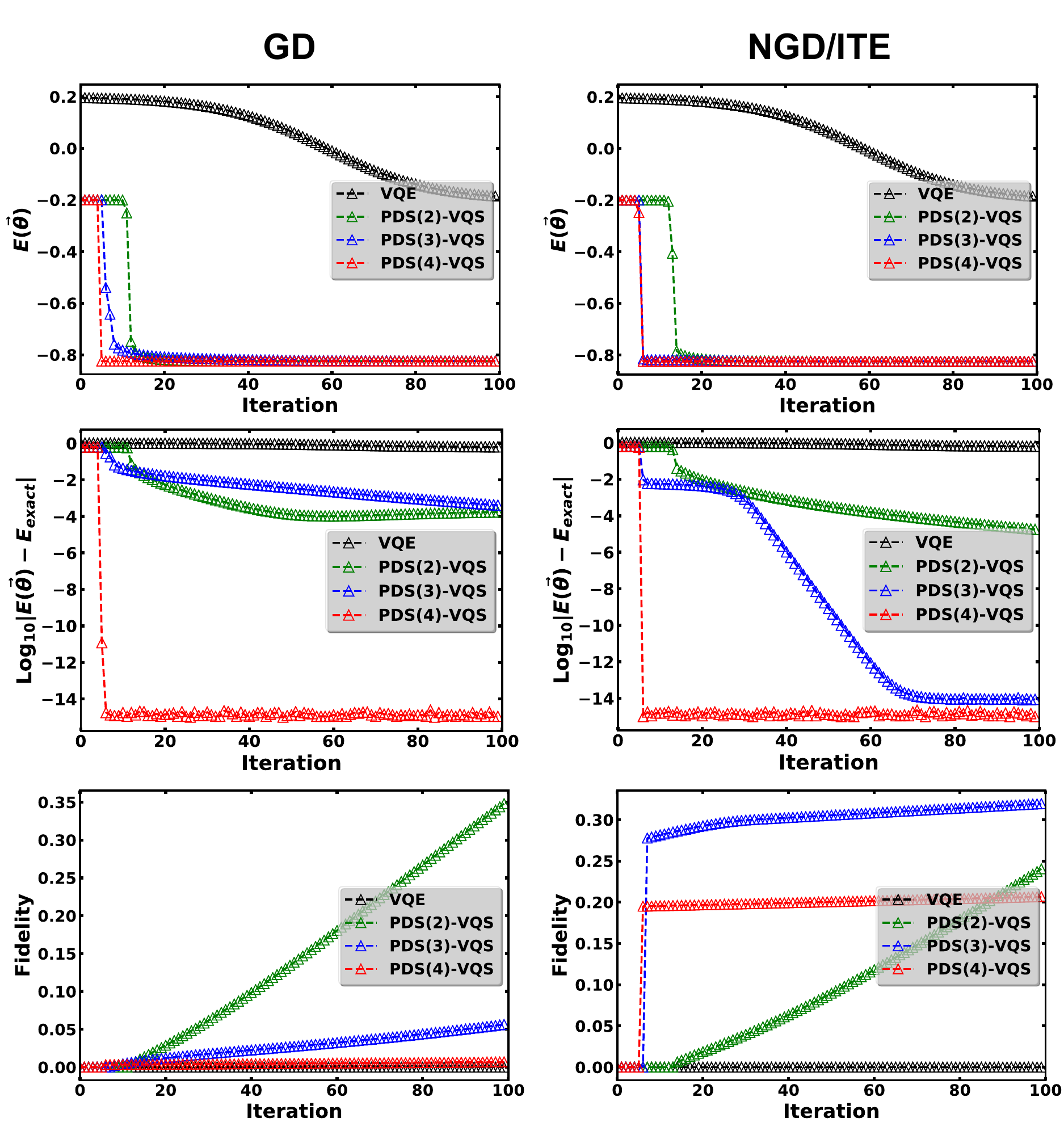}
    \caption{The computed ground state energy (top panels), energy deviation w.r.t. exact energy (middle panels), and fidelity of the trial state (bottom panels) of the H$_2$ molecule iterate in the conventional VQE and PDS($K$)-VQS ($K=2,3,4$) infrastructures employing gradient descent (left panels) and natural gradient descent/imaginary time evolution (right panels). The initial rotation is given by $\vec{\theta} = (7\pi/32,\pi/2,0,0)$. The step size $\eta=0.05$ in all the calculations.}
    \label{H2}
\end{figure}

Fig. \ref{H2} compares the VQE and PDS($K$)-VQS performances exploiting the above-mentioned ansatz to find the ground state energy of the H$_2$ Hamiltonian. As can be seen, starting from the given initial rotation, the VQE is unable to converge to the ground state energy within 100 iterations, but rather drops to an excited state energy ($-0.2$ a.u. in this case). Actually, it has been shown that,\cite{yamamoto2019natural} starting from the same initial rotation, the VQE needs to go through a ``plateau'' that resides at this energy value and spreads over $\sim$400 iterations before hitting the ground state energy ($\sim$$-0.8$ a.u. in this case) regardless of the employed Riemannian metric.

To achieve a higher level of accuracy (e.g. chemical accuracy $\|E(\vec{\theta})-E_{exact})\|<1.5\times10^{-3}$ a.u.), low order PDS($K$)-VQS typically needs more iterations than high order PDS($K$)-VQS. As shown in the middle left panel of Fig. \ref{H2}, by employing GD in the dynamics, it takes the PDS(4)-VQS $<$10 iterations to converge to the ground state energy with energy deviation being $<10^{-14}$ a.u. regardless of the employed Riemannian metric, while it takes the PDS(2)/PDS(3)-VQS almost 100 steps to bound the deviation to be $<10^{-3}$ a.u. Remarkably, the performance can be improved when GD is replaced by NGD/ITE in the PDS(2)/PDS(3)-VQS dynamics. In particular, within 80 iterations the PDS(3)-VQS employing NGD/ITE can converge to the accuracy level that is almost same as that of PDS(4)-VQS.

On the other hand, the quality of the trial wave function is more significantly improved in the low order PDS($K$)-VQS dynamics than in the high order PDS($K$)-VQS dynamics. For example, the fidelity of the trial wave function w.r.t the exact ground state gradually increases from almost zero to $\sim$0.35 within 100 iterations using PDS(2)-VQS regardless of the employed Riemannian metric, and this change is significantly steeper than the almost flat curves of PDS(3)/PDS(4)-VQS as shown at the bottom of Fig. \ref{H2}. However, in comparison to GD, employing NGD/ITE in the PDS(3)/PDS(4)-VQS quickly improves the fidelity of the trial wave function from $<$0.02 to 0.2$\sim$0.3 within 10 iterations. It is worth mentioning that since the fidelity of the trial wave function at the initial rotation is almost zero, both VQE and the static PDS($K$) ($K=2,3,4$) calculations alone cannot help identify the ground state energy in this case, which makes PDS($K$)-VQS a necessary and effective approach to target ground state energy and improve the trial wave function.

\textcolor{black}{Remarkably, the improvement of the trial wave function employing PDS($K$)-VQS approach might be limited. This can be seen from the flat fidelity curves of the trial state driven in the PDS(3/4)-VQS dynamics after first several iterations as shown at the bottom right of Fig. \ref{H2}.  This is due to the fact that the PDS($K$) formalism does not require the ansatz to sufficiently approximate the target state, while is still able to provide systematically improvable upper bounds of the expectation value of the target state by exploring the Krylov subspace. The benefit is the great simplification of the state preparation. The limitation is also obvious in that it sometimes would be challenging to further improve the quality of the trial state within the PDS($K$)-VQS framework if the energies were well converged already, which would then compromise the accuracy of the property calculations that usually requires a  sufficiently accurate description of the target wave function.}

\begin{figure}[h!]
    \includegraphics[width=\linewidth]{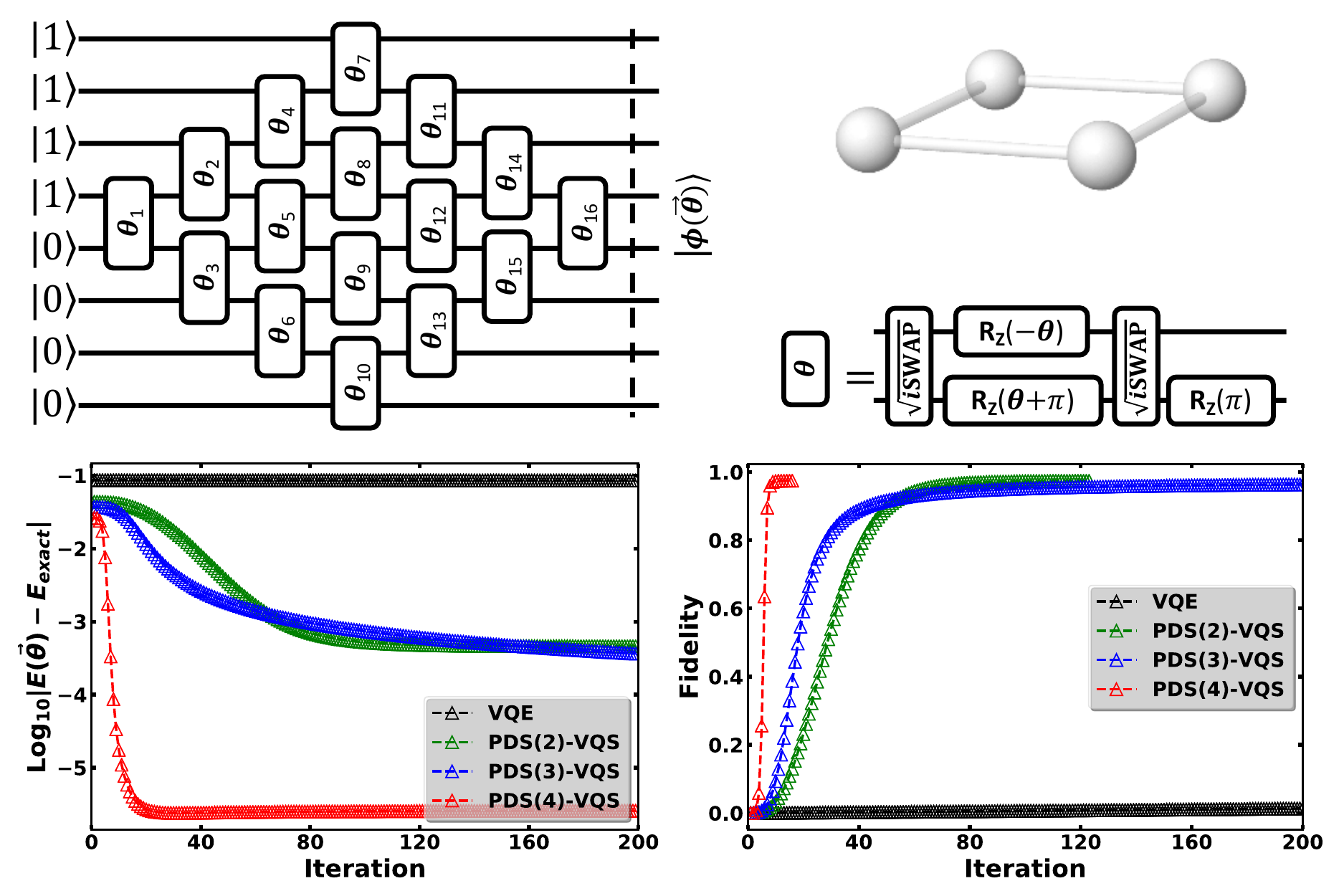}
    \caption{The performance of VQE and PDS($K$)-VQS ($K=2,3,4$) employing ordinary gradient descent (GD) to compute the ground state energy of a planar H$_4$ system with $R_{H-H}=2.0$ a.u. (Top left) The circuit used to generate the ansatz with 16 rotation parameters that is inspired by the basis rotation ansatz for a linear hydrogen chain in Ref. \cite{GoogleHF2020}. Here, we consider the planar H$_4$ system in 3-21G basis. The generated Hamiltonian acts on eight qubits, and considers an active space of four electrons in eight spin-orbitals. With this Hamiltonian, the ground state of the planar H$_4$ system is a triplet state with the exact energy $E_{\text{exact}}=-2.00591266$ a.u. (Bottom left) the deviations of the VQE and PDS($K$)-VQS energies and (bottom right) the fidelity change of the trial wave function w.r.t. true ground state during the PDS($K$)-VQS calculations. The initial values of all the rotations are set to 0.001. The step size $\eta=1.0$ in all the calculations.}
    \label{H4}
\end{figure}

We also test the proposed PDS($K$)-VQS approach for a slightly larger system, the planar H$_4$ system, where a 8-qubit circuit with 16 rotation parameters shown at the top of Fig. \ref{H4} is employed to prepare the trial wave function for finding the ground state energy. The state preparation circuit is inspired by the similar circuit that has been reported being successfully applied for preparing the trial state for \textcolor{black}{the Hartree Fock state of the } linear hydrogen chain systems.\cite{GoogleHF2020} 
For the planar H$_4$ system whose ground state is a triplet, the circuit with close-to-zero initial rotations would generate a trial state that is almost singlet, which makes the conventional VQE and the static PDS($K$) ($K=2,3,4$) simply fail. On the other hand, as shown at the bottom of Fig. \ref{H4}, the PDS($K$)-VQS ($K=2,3,4$) are capable of dealing with such a tough situation and again outperform.  As can be seen, within 200 iterations, PDS($K$)-VQS ($K=2,3,4$) are able to converge to the ground state energy well below chemical accuracy and improving the fidelity of the trial wave function to be $>$0.96. 
\textcolor{black}{It is worth noting that even though the converged rotations obtained from the PDS($K$)-VQS calculations generate a high fidelity state, the expectation value of the generated state is still $\sim0.02$ a.u. above the exact energy, and it then becomes challenging to further improve the fidelity employing the same circuit infrastructure through varying the rotations. Therefore, the circuit used here might not be sufficient for preparing true ground state in practice if higher fidelity is desired. We here intentionally employ the circuit to artificially generate an extreme challenging case to show the performance difference between conventional VQE and PDS($K$)-VQS approaches.}

\section{Discussion}

\textcolor{black}{From Section III, it has been seen that the PDS($K$)-VQS approach bears the potential of speeding up the iterations in comparison with the conventional VQE approach. However, it is worth noting that the measurement effort of evaluating $\langle H^n \rangle$'s ($n>1$) and their derivatives are usually more expensive than that of $\langle H \rangle$ and its derivative, and the actual cost saving will therefore be compromised.}

\textcolor{black}{To have a close look at the measurement of the $\langle H^n \rangle$ (and its impact on the total cost), we employ the following metric  to give an estimate for the number of measurements, $M$,\cite{gonthier2020identifying,Rubin2018Hybrid,PhysRevA.92.042303}
\begin{align}
   M = \Bigg( \frac{\sum_G\sqrt{\sum_{i,j,\in G}h_i h_j \text{cov}\big(P_i,P_j\big)}}{\epsilon}\Bigg)^2
   , \label{eq:Measurement}
\end{align}
where $\epsilon$ is the desired precision and $h_i$'s and $P_i$'s are the coefficients and Pauli strings representing a moment (i.e. $H^n=\sum_i h_i P_i$) and having been partitioned into certain groups, $G$'s, in which simultaneous measurement can be performed. $\text{cov}\big(P_i,P_j\big)$ is the covariance between two Pauli strings bounded by
\begin{align}
    \text{cov}\big(P_i,P_j\big) \leq | \sqrt{\text{var}(P_i) \cdot \text{var}(P_j)} |
\end{align}
with the variance being computed from $\text{var}(P_i) = 1 - \langle P_i\rangle^2$. Here, we assume the covariances between different Pauli strings to be zero for the brevity of the discussion. We can apply the above metric to, for example, estimate the number of measurements of $H^n$ ($n=1,2,3$) required by the PDS(2)-VQS calculation for the complete active space (4 electrons, 4 spin-orbitals) of the planar H$_4$ system. Given $\epsilon\sim0.5$mHartree, since $H^n$ ($n=1,2,3$) can be generated from at most $\sim3700$ Pauli strings, the estimated number of measurements needs to be done is $\sim4.8\times10^{9}$, which is one order of magnitude higher than that for $\langle H\rangle$ ($\sim1.2\times10^8$). Thus, given the same trial state in this H$_4$ case, if the number of conventional VQE iterations is no more than one order of magnitude larger than that of the PDS(2)-VQS iterations, VQE would outperform PDS(2)-VQS in terms of total number of measurements, and PDS(2)-VQS outperforms otherwise. It is worth mentioning that, during the PDS($K$)-VQS process for the ground state and energy, the excited state energies can also be estimated directly from the higher roots of the polynomial (12) without any additional measurement (although accurate excited state energies would require higher order PDS($K$)-VQS calculations). In contrast, the conventional VQE would need distinct trial states, and thus different measurements, for targeting different states.}

\textcolor{black}{Generally speaking, as long as the relatively large number of measurements of the Pauli strings becomes manageable, the PDS($K$)-VQS approach can be potentially applied for targeting the exact solutions for the system sizes that are not classically tractable, in particular for the systems whose true ground and excited states we have little knowledge of, or are challenging to obtain classically. To reduce the measurement demand, typical strategy is to partition the Pauli strings (that contribute to the moments) into commuting subsets that follow a certain rule, e.g. qubit-wise commutativity (QWC)\cite{kandala2017hardware,mcclean2016theory}, general commutativity\cite{gokhale2020,Izmaylov2020}, unitary partitioning\cite{Izmaylov2019}, and/or Fermionic basis rotation grouping\cite{Huggins_2021} to name a few. The applications of these commuting rules to the single Hamiltonian have shown that, at a cost of introducing additional one-/multi-qubit unitary transformation before the measurement, the total number of required measurements can be significantly reduced from $\mathcal{O}(N^4)$ to $\mathcal{O}(N^{2\sim3})$, or for simpler cases even $\mathcal{O}(N)$. For higher order moments, as we mentioned in the method section, early study of applying QWC bases to Heisenberg models represented by up to 36 qubits exhibits a sub-linear scaling of the number of measurements in the number of the qubits (Ref. \cite{Vallury2020quantumcomputed}), which then leads us to expect similar scaling behaviors of the number of required measurements for evaluating moments for molecular systems. Beside exploring the commutativity of Pauli strings, other approaches including the linear combinations of unitary operators (LCU) technique\cite{Childs2012}, direct block-encoding\cite{gilyen2019quantum,berry2015Simulating}, and quantum power methods\cite{Seki2021quantum} might also be worth studying for reducing the number of measurements at the cost of circuit depth. In the light of that, we plan to perform a comprehensive benchmark as a follow-up work.}

\textcolor{black}{Since the PDS($K$)-VQS formalism involves solving linear system of equations and polynomial, there is a concern of numerical instability when applying the PDS($K$)-VQS approach in optimization. Theoretically, the numerical instability of the PDS($K$)-VQS approach might come from two sources, (a) the singularity and ill-conditioning of the matrix $M$ in Eq. (\ref{lineq1}) that might consist of high order moments, and (b) the singularity of the Riemannian metric ($\mathcal{R}$) used in the dynamics (\ref{dynamics}). In particular, the singularity of matrix $M$ can be easily observed if the trial vector becomes very close to the exact wave function ($\rm det |M| = 0$ if we replace the trial vector with exact vector). Numerically, the singularity problem can be avoided by adding a small positive number (e.g. $10^{-6}$) to the eigenvalue of the matrix $M$ or $\mathcal{R}$ via singular value decomposition (SVD). However, it is worthing noting that adding small perturbation to $M$ might violate the variationality of the PDS approach, and would not be recommended to use if the strict upper-bounds to the true energy are concerned. The ill-conditioning of matrix $M$ could occur in the high order PDS calculations, where high order moments could make the condition number of matrix $M$ very large. Thus, from the practical point of view, due to the potentially larger number of measurements and ill-conditioning arising from high Hamiltonian powers, lower oder PDS($K$)-VQS approaches are usually more feasible.}

\begin{figure}[h!]
    \includegraphics[width=\linewidth]{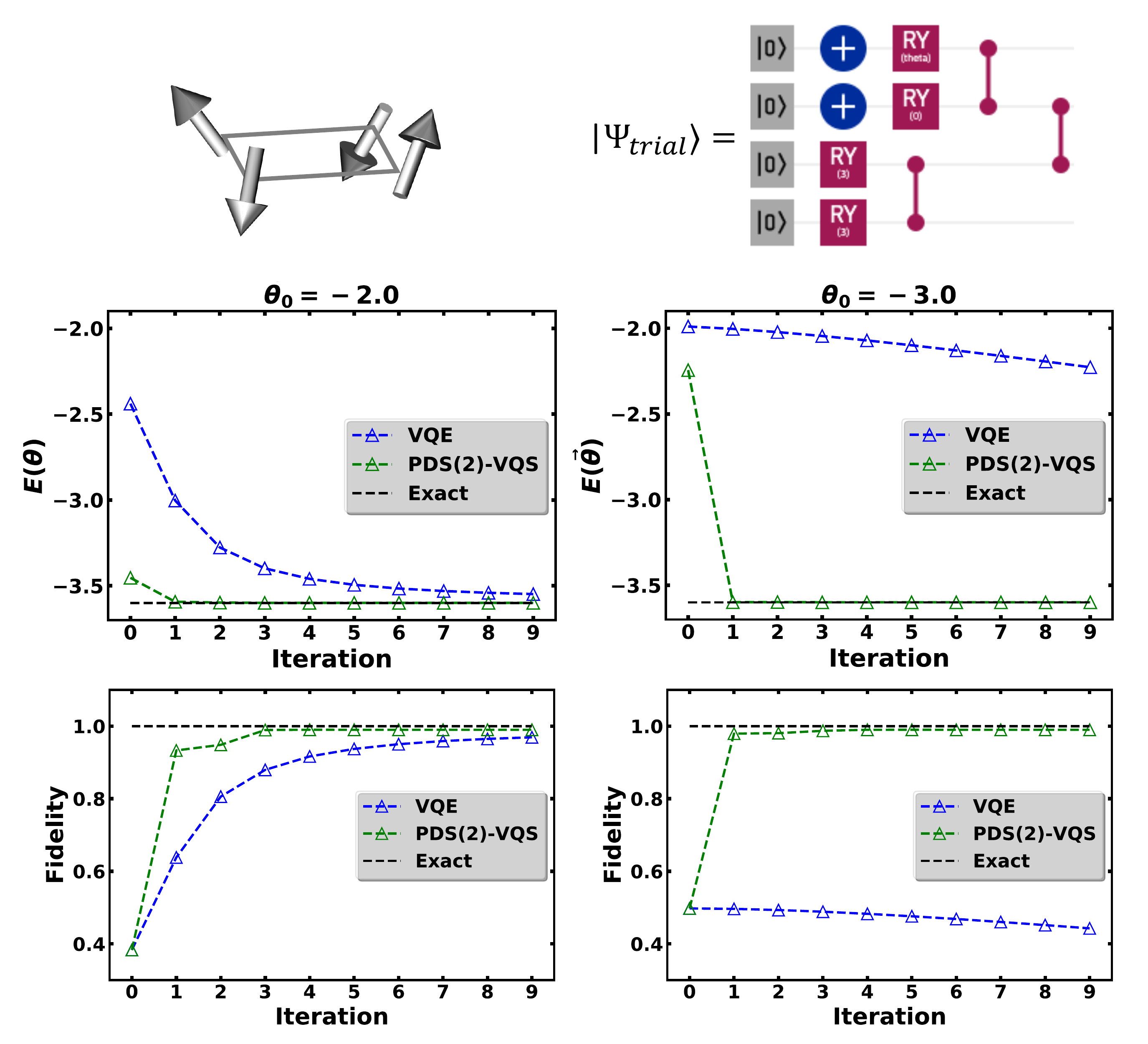}
    \caption{\textcolor{black}{The performance of VQE and PDS(2)-VQS employing ordinary gradient descent (GD) to compute the ground state energy of the four-site 2D Heisenberg model. (Top right) The circuit employed to generate the trial vector, where only the first rotation in RY gate is treated as a variational parameter $\theta$, and other three rotations are fixed to $(0,3,3)$. Two initial rotations $\theta_0=-2.0$ and $\theta_0=-3.0$ are chosen for performance comparison. The exact ground state energy of the 2D Heisenberg model is $E_{\text{exact}}=-3.6$ a.u. (Center) The VQE and PDS(2)-VQS energies and (bottom) the corresponding fidelity changes of the trial vectors w.r.t. true ground state in the first ten iterations in the conventional VQE and PDS(2)-VQS noise-free calculations. The step size $\eta=1.0/\text{Iteration}$ in all the calculations.}}
    \label{Heisenberg_ideal}
\end{figure}

\begin{figure}[h!]
    \includegraphics[width=\linewidth]{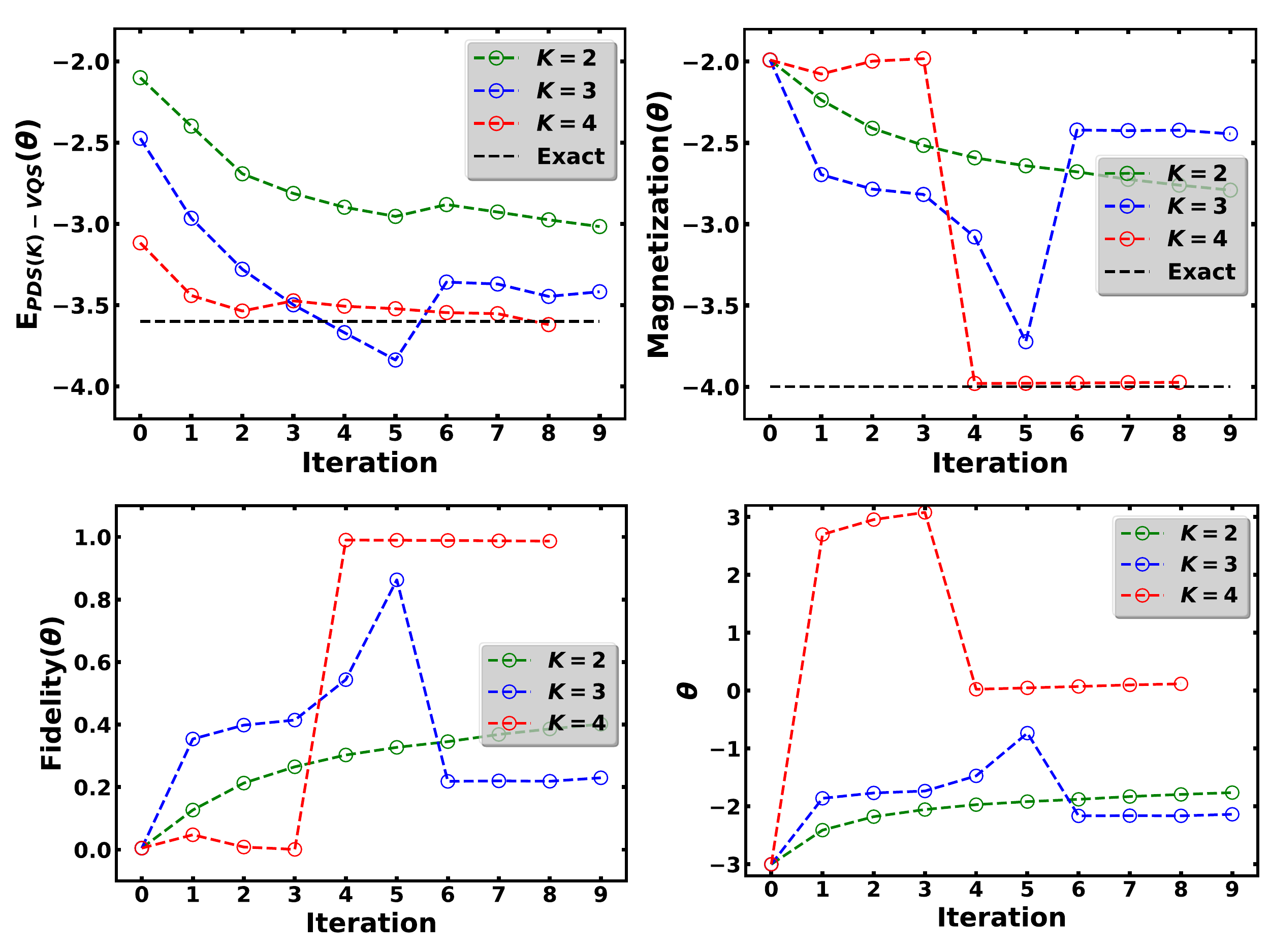}
    \caption{\textcolor{black}{The computed ground state energy (top left) and magnetization (top right) of the four-site 2D Heisenberg model and the corresponding changes of the fidelity (bottom left) and variational parameter $\theta$ (bottom right) in the first ten PDS($K$)-VQS ($K=2,3,4$) iterations running on IBM Toronto quantum hardware. The physical setup, error sources, and computed expectation values of Hamiltonian moments (up to $\langle H^7 \rangle$) and the associated standard deviations are shown in Fig. \ref{Heisenberg_appendix}. In all the calculations ordinary gradient descent (GD) is employed.  The initial rotation $\vec{\theta}_0=-3.0$. The exact ground state energy and magnetization of the 2D Heisenberg model are $E_{\text{exact}}=-3.6$ a.u. and $\sum_i\langle\sigma_{z_i}\rangle=-4.0$ a.u., respectively. The step size $\eta=1.0/\text{Iteration}$ in all the calculations.}}
    \label{Heisenberg_ibm}
\end{figure}

\textcolor{black}{Ultimately, one would be concerned about how the PDS($K$)-VQS applies to general models and how it performs on the real quantum hardware subject to the device noise. To address these concerns and explore the potential of the PDS($K$)-VQS approach, we have started to launch the PDS($K$)-VQS calculations for more general Hamiltonians on both simulator and the real quantum hardware. Figs. \ref{Heisenberg_ideal} and \ref{Heisenberg_ibm} exhibit some preliminary results for a four-site 2D Heisenberg model with external magnetic field, $H = J\sum_{\langle ij \rangle} \big( X_iX_j + Y_iY_j + Z_iZ_j\big) + B\sum_i Z_i$ with $J/B=0.1$. The simple circuit employed for the state preparation in both VQE and PDS($K$)-VQS simulations is shown in Fig. \ref{Heisenberg_ideal}, where, for the brevity of our discussion, we only treat one rotation in the state preparation as the variational parameter, and fix all other three rotations. As can be seen from the noise-free simulations in Fig. \ref{Heisenberg_ideal}, the PDS(2)-VQS results quickly converge within five iterations achieving $\sim0.99$ fidelity, while the performance of VQE exhibits strong dependence on the initial rotation (for $\vec{\theta}_0=-2.0$,  the conventional VQE is able to converge in 10 iterations with $\Delta E<0.05$ a.u. and Fidelity $\sim0.97$).
When running the PDS($K$)-VQS simulations for the same model on the IBM Toronto quantum hardware, as shown in Fig. \ref{Heisenberg_ibm}, in comparison to the ideal curves, the PDS(2/3)-VQS optimization curves on the real hardware significantly slows down, and deviate from the exact solutions due to the error from the real machine. However, if we increase the PDS order to perform PDS(4)-VQS calculations, the accuracy of the results systematically improves. For example, in the PDS(4)-VQS approach both the computed ground state energy and the trial state (and thus the magnetization) converge within 10 iterations being very close to the exact solutions.}

\section{Conclusion}
In summary, we propose a new variational quantum solver that employs the PDS energy gradient.  In comparison with the usual VQE, the PDS($K$)-VQS helps identify an upper bound energy surface for the ground state, and thus frees the dynamics from being trapped at local minima that refer to non-ground states. In comparison with the static PDS($K$) expansions, the PDS($K$)-VQS guides the rotation of the trial wave function of modest quality, and is able to achieve high accuracy at the expense of low order PDS($K$) expansions.  We have demonstrated the capability of the PDS($K$)-VQS approach at finding the ground state and its energy for toy models, H$_2$ molecule, and strongly correlated planar H$_4$ system in some challenging situations. In all the case studies, the PDS($K$)-VQS outperforms the standalone VQE and static PDS($K$) calculations in terms of efficiency even at the lowest order. \textcolor{black}{We also discussed the limitations of the PDS($K$)-VQS approach at the current stage. In particular, the PDS($K$)-VQS approach may suffer from large amount of measurements for large systems, which can nevertheless be reduced at the cost of circuit depth by working together with some measurement reduction methods. Finally, we have started to launch PDS($K$)-VQS simulations for more general Hamiltonians on IBM quantum hardware. Preliminary results for Heisenberg model indicate that higher order PDS($K$)-VQS approach exhibits better noise-resistance than the lower order ones. The discussed approach can be extended to any variational formulation based on the utilization of  
$\langle H^n \rangle$ moments (e.g. Krylov subspace algorithms).}

\onecolumn\newpage

\begin{figure*}[h!]
    \includegraphics[width=\textwidth]{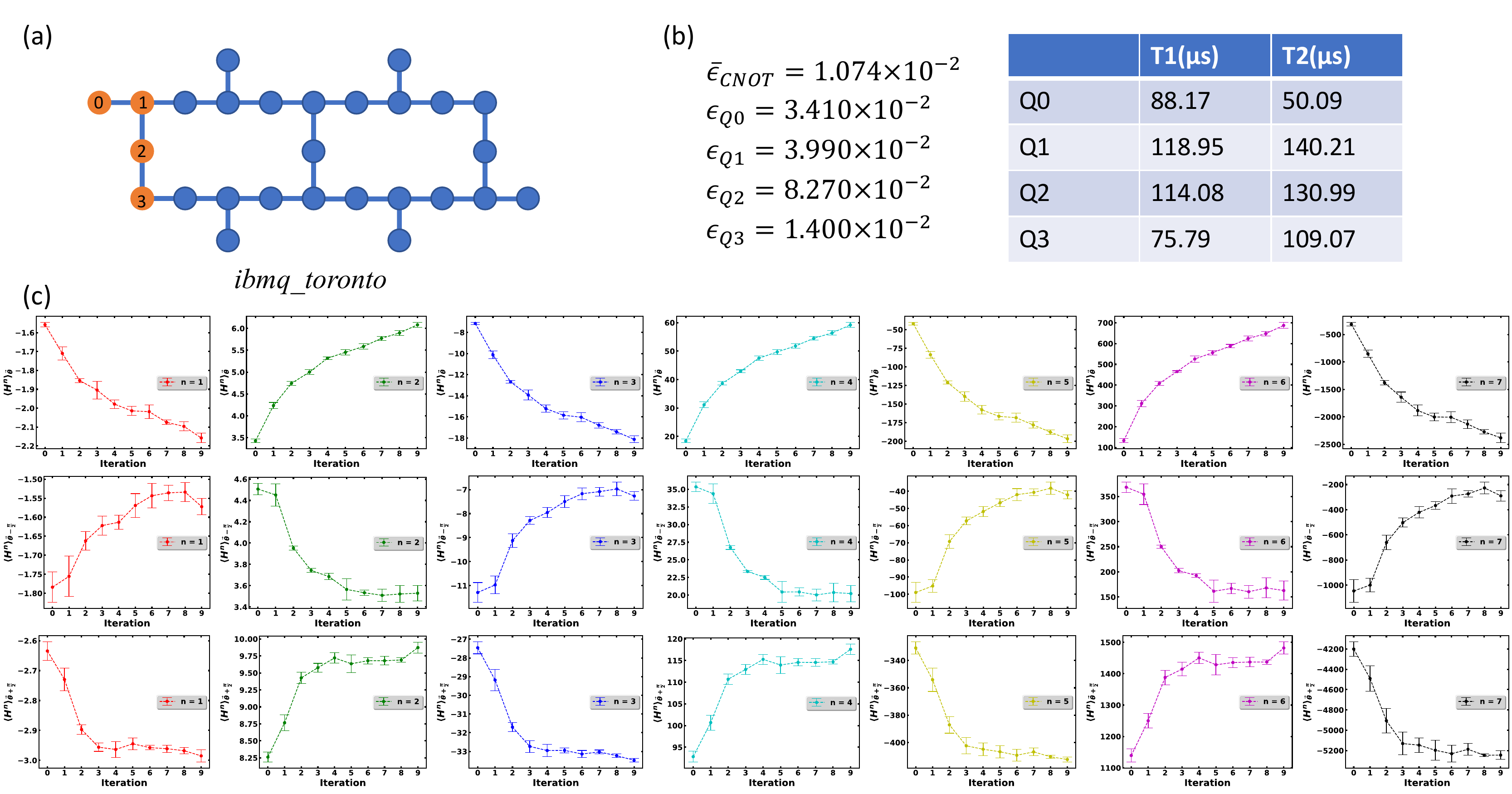}
    \caption{(a) Quantum processor device map for \textit{ibmq\_toronto} showing the four qubits (Q$_n$, $n=0-3$) used in the present computation. (b) Average CNOT error, 1-qubit readout assignment error, and thermal relaxation time constant (T1) and dephasing time constant (T2) in the four qubits used in the present computation. (c) The expectation values of the Hamiltonian moments, $\langle H^n \rangle$ ($n=1-7$), assembled from the measurements of the expectation values of  21 QWC bases for four-site 2D Heisenberg model $H = J\sum_{\langle ij \rangle} \big( X_iX_j + Y_iY_j + Z_iZ_j\big) + B\sum_i Z_i$ with $J/B=0.1$. The data points correspond to mean value from the calculations on IBM Quantum processor \textit{ibmq\_toronto} with statistical error bars corresponding to $5\times8192$ shots (per point). The trial state is constructed using the circuit given in Fig. \ref{Heisenberg_ideal} with initial rotation $\theta_0=-3.0$.}
    \label{Heisenberg_appendix}
\end{figure*}

\twocolumn
\section{Acknowledgement}
B. P. and K. K. were supported by  the ``Embedding QC into Many-body Frameworks for Strongly Correlated  Molecular and Materials Systems'' project, which is funded by the U.S. Department of Energy, Office of Science, Office of Basic Energy Sciences (BES), the Division of Chemical Sciences, Geosciences, and Biosciences. B. P. and K. K. acknowledge the use of the IBMQ for this work. The views expressed are those of the authors and do not reflect the official policy or position of IBM or the IBMQ team.

\section{Data Availability}
The data that support the findings of this study are available from the corresponding author upon reasonable request.

\bibliographystyle{plainnat}
\bibliography{ref}

\end{document}